\documentclass{caosp302}

\usepackage{graphicx}

\articleNo{123}
\pubyear{2008}
\volume{35}
\volnumber{3}
\firstpage{1}
\received{}
\accepted{}

\begin{document}

\hauthor{M.\,Gebran, R.\,Monier and O.\,Richard}

\title{Abundances determination of A, Am and F stars in the Pleiades and Coma Berenices clusters}

\author{
       M.\,Gebran \inst{1,}  
      \and 
        R.\,Monier \inst{2}   
      \and 
        O.\,Richard \inst{1}
       }

\institute{
           Groupe de Recherche en Astronomie et Astrophysique du Languedoc,UMR 5024, Universit\'e Montpellier II, Place Eug\`ene Bataillon, 34095 Montpellier, France, \email{gebran@graal.univ-montp2.fr}
         \and 
           Laboratoire Universitaire d'Astrophysique de Nice, UMR 6525, Universit\'e de Nice - Sophia Antipolis, Parc Valrose, 06108 Nice Cedex 2, France. 
                  }

\date{March 8, 2003}

\maketitle

\begin{abstract}
Abundances of 18 chemical elements have been derived for 27 A/Am and 16 F stars members of the Pleiades and Coma Berenices open clusters. We have specifically computed, with the Montr\'eal code, a series of evolutionary models for two A stars members of these two clusters. None of the models reproduces entirely the overall shape of the abundances patterns. The inclusion of competing processes such as rotational mixing in the radiative zones of these stars seems necessary to improve the agreement between observed and predicted abundances patterns.

\end{abstract}

\section{Introduction}
Stars in open clusters have the same initial chemical composition and age. They are very useful to test the predictions of evolutionary models. Models for F and A stars have been already claculated by Turcotte et al. (1998) and Richer et al. (2000) respectively. The main thrust of this paper is to compare the derived abundances with new self-consistent models calculated with the Montr\'eal code.

\section{Method}
Our observing sample consists of 22 A/F stars members of the Coma Berenices cluster and 21 A/F stars of the Pleiades open cluster. These stars were observed using ELODIE and SOPHIE \'echelle spectrographs at the Observatoire de Haute-Provence (OHP). For each star, effective temperature ($T_{\rm{eff}}$) and surface gravitie ($\log g$) were determined using Napiwotzki et al's (1993) UVBYBETA code based on Str\"omgren's $uvby\beta$ photometric indices. LTE model atmospheres were computed using Kurucz's ATLAS9 code (Castelli \& Kurucz 2003). The linelist was constructed from Kurucz's gfall.dat (http://Kurucz.harvard.edu). The oscillator strengths were checked against more accurate laboratory determinations when available. Synthetic spectra were computed using Takeda's (1995) iterative procedure. The microturbulent velocities, rotational velocities and the abundances of 18 elements were determined by fitting the line profiles of the observed spectrum.

\section{Results}
\begin{figure}[h!]
\begin{tabular}{cccccc}
\centering
&&&&
\includegraphics[scale=0.26]{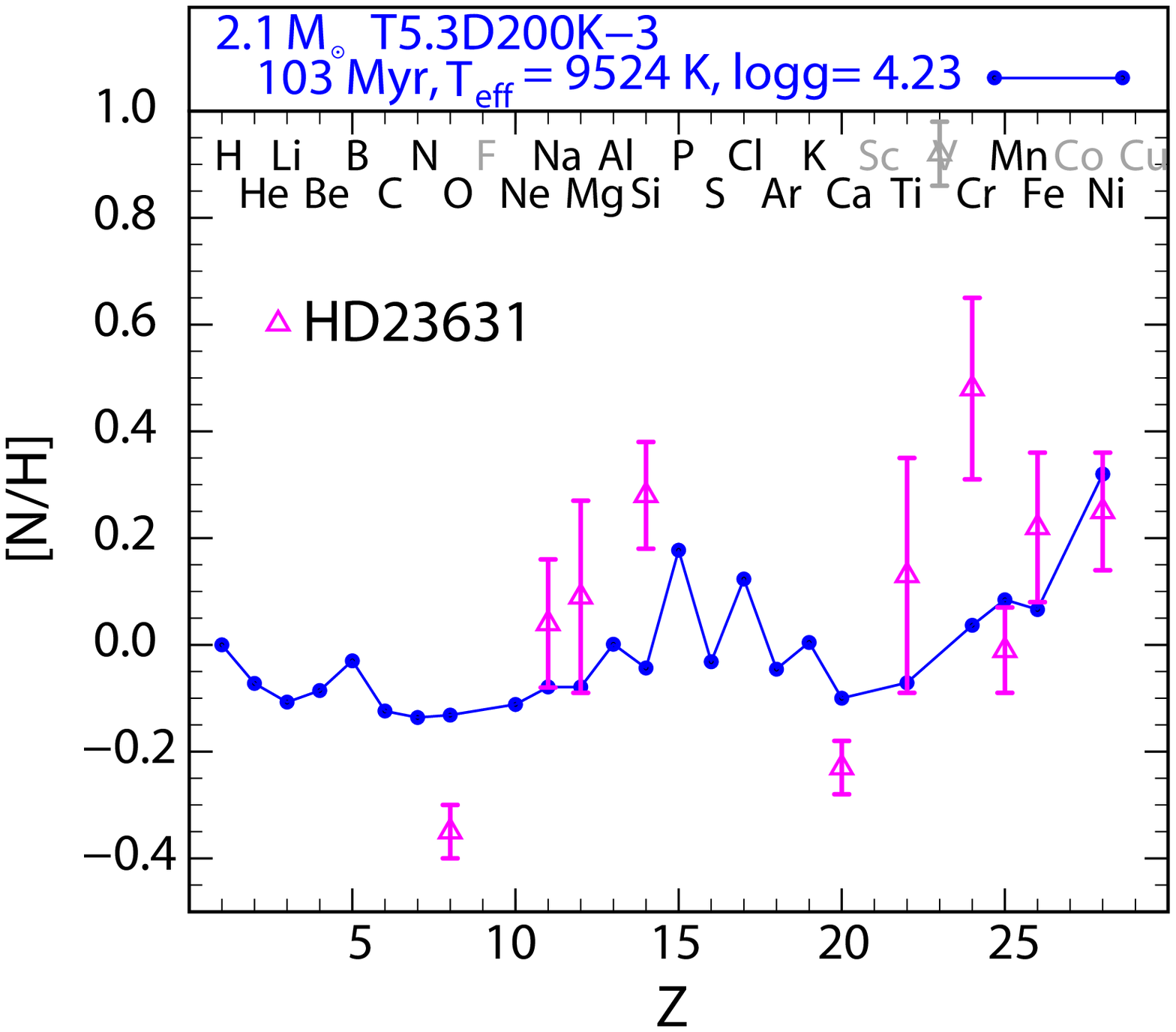}&
\includegraphics[scale=0.26]{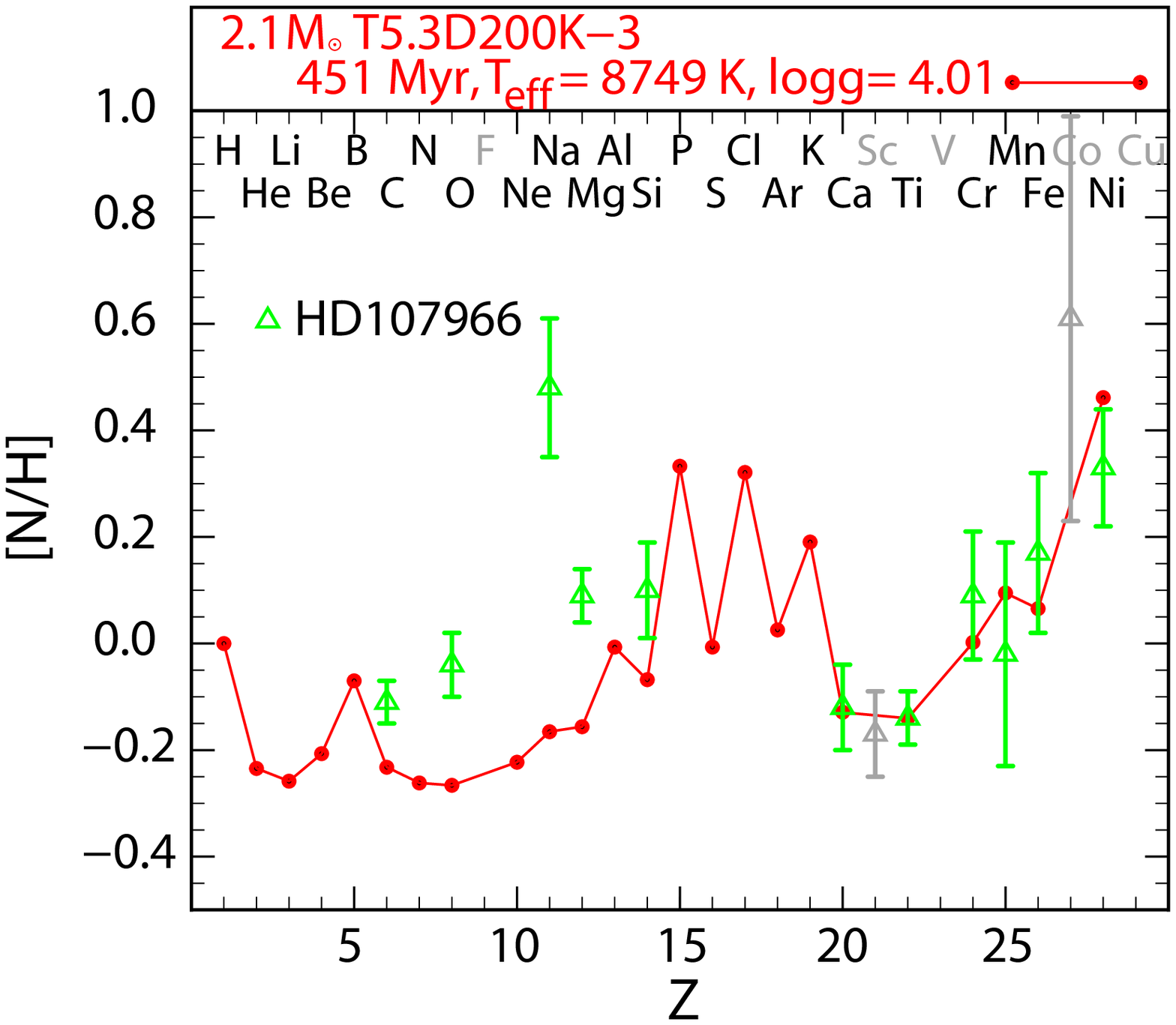}
\end{tabular}
\caption{Predicted surface abundances at 100 Myr for HD23631 (left panel) and at 459 Myr for HD107966 (right panel). Triangular symbols represent the observed abundances with representative error bars.}
\label{modele-evolution}
\end{figure}

For A stars, we found large star-to-star variations in the abundances of most the elements. Evolutionary models that include the effects of atomic and turbulent diffusion were calculated for 2 A stars, HD23631 member of Pleiades cluster and HD107966 member of Coma Berenices cluster. These clusters have very similar metallicity, therefore we can follow the evolution of the surface chemical composition of a star at the age of the Pleiades (100 Myr) to the age of Coma Berenices (450 Myr). After a series of tests, the adopted model was the T5.3D200k-3 (see Richer et al. 2000 for details). Most of the iron peak elements are well fitted by the model (figure \ref{modele-evolution}). A few discrepancies exist for light elements, Na, Mg and Si. Part of these discrepancies may arise from non-LTE effects. However, the inclusion of competing processes such as rotational mixing in the radiative zone (or mass loss) should help reproduce the observed patterns.

\end{document}